\documentclass[aps,prd,preprint,showpacs,amsmath,amssymb]{revtex4}
\usepackage{epsfig}
\newlength{\www}

\newcommand{\be}{\begin{equation}}
\newcommand{\ee}{\end{equation}}
\newcommand{\ba}{\begin{eqnarray}}
\newcommand{\ea}{\end{eqnarray}}

\begin{document}

\title{\vspace{1cm}
Yukawa enhanced electroweak corrections\\
at high energy in the MSSM
}

\author{M. Beccaria$^{a, b}$, E. Mirabella$^a$}

\affiliation{\small
$^a$ Dipartimento di Fisica, Universit\`a di Lecce \\
Via Arnesano, 73100 Lecce, Italy\\
\vspace{0.2cm}
$^b$ INFN, Sezione di Lecce\\
\vspace{0.2cm}
}

\begin{abstract}
We consider the electroweak radiative corrections in the MSSM and study the purely Yukawa 
contributions ${\cal O}(\alpha^L y_t^{2p} y_b^{2q})$, $p+q=L$, where $y_{t,b}$ are the top and bottom quark Yukawa couplings.
We show that these corrections can be computed in a gaugeless limit of the MSSM where they are under Renormalization Group control.
As an application, we present explicit results for various ILC and LHC processes valid at all orders in the loop expansion and at leading 
order in the large logarithms that arise at high energy.
\end{abstract}


\maketitle

\section{Introduction}
\label{sec:intro}

The Minimal Supersymmetric Standard Model (MSSM)~\cite{Martin} is one of the 
major theoretical laboratories
where physics beyond the Standard Model can be tested at quantitative level. 
Radiative corrections in the MSSM are under control at all orders~\cite{Hollik:2002mv}, at least 
in processes where a perturbative treatment is adequate. Of course, 
the large number of free parameters of the MSSM remains a main practical problem. The comparison between theory and experiment is difficult to perform
in a model independent way and with no specific assumptions.

Some simplification in the treatment of radiative corrections could be quite
useful to disentangle individual and different physical mechanisms. 
High energy expansions of  virtual effects are a natural candidate in such direction~\cite{LSE}.
Indeed, both at the Large Hadron Collider (LHC) and at the International Linear Collider (ILC) the experimental 
setup will be characterized by a large available energy  compared to the typical mass scales, at least in 
favorable (but not at all exotic) scenarios.

High energy expansions are based on the simple observation that asymptotic scattering amplitudes 
are dominated  by large radiative corrections growing with energy. For brevity, we shall indicate collectively all these contributions
as {\em Sudakov logarithms} (SL), although only a part of them deserves this terminology.
A detailed quantitative analysis of the Sudakov effects is by now very well understood 
both in the Standard Model~\cite{SudakovSM:Reviews} and in the MSSM~\cite{SudakovMSSM:Reviews}.

At one loop and in $2\to 2$ processes, SL can be classified in four very different classes~\cite{Beccaria:2003yn}. 
We concentrate on the electroweak sector and discuss the four 
categories emphasizing in particular the status of their theoretical control.
The detailed application of this formalism to realistic physical problems can be found 
in~\cite{SudakovMSSM:ILC} for ILC processes and in~\cite{SudakovMSSM:LHC1,SudakovMSSM:LHC2}
for LHC processes.

The first class is that of {\em Universal SL} which are infrared contributions.
The $W^\pm$ and $Z^0$ gauge boson masses act as infrared regulators in the loop diagrams and we can observe
enhanced corrections at high energy with leading form at one loop $\sim\alpha \log^2(s/M_W^2)$, where $\sqrt{s}$ is the center of mass energy and 
$\alpha = e^2/4\pi$. These mass
singularities are called universal because they receive a contribution from each external particle (initial or final)
independently of the diffusion details. Also, the weight of the squared logarithm is a simple 
combination of the gauge group quantum number of the external particles. At $L$ loops, the leading 
contributions of this kind are $\sim\alpha^L \log^{2L}(s/M_W^2)$ with subleading corrections involving
all powers smaller than $2L$. 
The universal SL are the most relevant from the numerical point of view. For this
reason there have been large efforts to compute them at higher orders or (working at fixed order in $\alpha$)
to extend the calculation to additional subleading terms. The theoretical control of the universal SL
is coded in the hard evolution equations. Detailed applications at the two loop level
in the Standard Model can be found in~\cite{HardEvolution}.
Explicit diagrammatic calculations have also been accomplished~\cite{Denner:2003wi}.

In the MSSM, the existing calculations are at the one loop level, with the exception of a certain
number of resummation proposals valid at all orders in the perturbative expansion and at NLO 
in the logarithms~\cite{YukawaResummation}.

The next class we consider is that of {\em Renormalization Group} (RG) SL. These are complementary to the universal ones.
Indeed, RG-SL are effects with an ultraviolet origin and arise as a consequence of the running of the 
coupling constants. Being a short distance effect 
they are expected to be independent on the long distance physics.
At $L$ loops they have the leading behavior $\sim\alpha^L\log^L(s/M_W^2)$. 
Since there is only one logarithm per loop and not two as in the universal case, 
the RG Sudakov contributions are NLO at one loop, NNLO at two loop and so on. 
Thus, although they are completely under control from the theoretical point of view, they are not expected to 
be the dominant contribution in any reasonable limit.

The third class is composed of the {\em Angular} SL. At one loop, 
they are contributions coming from Standard Model box diagrams and with  asymptotic behavior
 $\sim\alpha \log(s/M_W^2) \log|x/s|$ where $x=t$ or $u$, the Mandelstam variables involving
the scattering angle. The Standard Model contributions have been investigated at two loops~\cite{Denner:2003wi}.
They do not receive supersymmetric corrections and we shall regard them as known without further discussion.

The fourth and last class is the most interesting for our purposes and consists of the {\em Yukawa} SL.
These contributions are present only when heavy quarks or their SUSY partners are produced in the final state
(or belong to the initial state, at LHC). More precisely, Yukawa SL are effects with characteristic factors 
$m_{t,b}/M_W$ where $m_{t,b}$ are the top and bottom quark masses~\cite{SudakovMSSM:YukawaEffects}.
These factors are of course Yukawa couplings in disguise. 
Again, at leading order, we have one logarithm for each loop. In the Standard Model, Yukawa SL are 
numerically non negligible, but not so interesting for the same reasons we mentioned about RG-SL.
However, in the MSSM the Yukawa couplings can be enhanced in certain regions of the parameter space 
and Yukawa SL permit interesting phenomenological analyses. These regions
are characterized by large values of the mixing parameter $\tan\beta$. This important parameter 
is singled out in the Yukawa SL weights and it is possible to fix bounds on it from the Sudakov
expansion of the observables of  several processes~\cite{SudakovMSSM:YukawaEffects}. 

The theoretical control over Yukawa SL is not very satisfactory as it stands. 
In the Standard Model, they have been often neglected by working in processes involving only light fermions~\cite{HardEvolution}. In the MSSM, 
where the top quark physics programme is very promising, the existing calculations are limited to the one loop level.

On the other hand, the one loop analysis suggests that a more detailed investigation should be possible because of some 
remarkable features~\cite{SudakovMSSM:YukawaEffects,Beccaria:2005xq}.
First, the Yukawa SL turn out to be correction factors associated to the external particles
and not to the specific details of each process. This is similar to what happened with the universal SL.
Second, there is only one logarithm per loop and this seems to be the marker of a ultraviolet effect.
{\em A posteriori}, it is strange that there are RG-SL related to the gauge coupling running, but nothing 
analogous in the Yukawa sector where additional independent couplings are present. 

In this paper, we fill these gaps and explain the above features. We show 
that Yukawa SL are indeed a short distance effect which is governed by Renormalization Group equations 
in the gaugeless limit of the MSSM, {\em i.e.} in the Yukawa sector. 
Contrary to the RG-SL, we have motivations to pursue a calculation of Yukawa SL at higher order. We
anticipated the reason and we repeat it here to emphasize its importance. There are regions in the MSSM
parameter space where the Yukawa couplings can be enhanced. If we are in such a region, the Yukawa SL
can give large corrections at one loop, comparable to universal SL, although they are subleading in a formal 
logarithmic expansion. For the purposes of a high precision measurement, the determination of the next
perturbative SL contributions acquire therefore a substantial relevance.

\section{Yukawa enhanced electroweak corrections in the MSSM at high energy}
\label{sec:general}

\begin{figure}[htb]
\vskip 0.7cm
\begin{center}
\leavevmode
\psfig{file=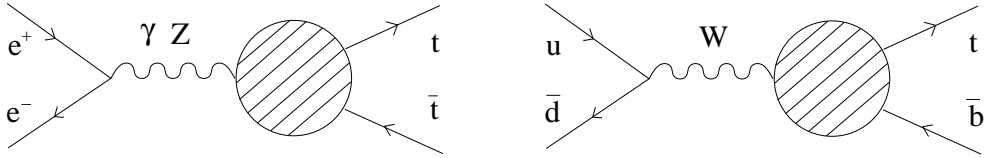,width=13cm,angle=0}
\vspace{0.1cm}
\caption{Two sample processes relevant for ILC and LHC respectively. In both processes the leading Yukawa SL
are confined in the dashed circles representing the final dressed interaction vertex.}
\label{fig:diagrams}
\end{center}
\end{figure}

The detailed structure of Yukawa enhanced Sudakov corrections in the MSSM is illustrated by two sample processes
shown in Fig.~(\ref{fig:diagrams}).
\noindent
The first example, $e^+e^-\to t\overline{t}$, is a typical $s$-channel neutral process and describes top quark pair production at ILC.
The second example is instead a charged process, $u\overline{d}\to t\overline{b}$, and is one of the 
three partonic processes which permit single top quark production at LHC~\cite{CERNYB}. In the Feynman diagrams we have 
drawn the final vertex by means of a dashed circle standing for a particular set of radiative corrections. These are precisely
the electroweak virtual exchanges that give rise to \underline{purely} Yukawa SL, {\em i.e.} all contributions
that at the $L$ loop level and at high energy correct the tree level amplitude by factors that take the form 
\be
\label{eq:yuk}
\alpha^L \left(\frac{m_t^2}{M_W^2}\right)^p\left(\frac{m_b^2}{M_W^2}\right)^q \log^r\frac{s}{M^2},\quad p+q = L, 
\ee
where $m_{t,b}$ are the top and bottom masses, $s$ is the squared c.m. energy, and $M$ is a typical process mass scale. 
We emphasize that there are also SL with factors like above, but with $p+q<L$. We shall not discuss these contributions
which mix the infrared and ultraviolet structure of the MSSM.
In particular, the papers~\cite{YukawaResummation}  resum SL at NLO order (that is including
terms $\sim \log^{2L-1} s$) and determine at $L>1$ the contribution of the form 
Eq.~(\ref{eq:yuk}) with $p+q=1$ and $r=2L-1$. Our calculation extends their result by independent additional terms.
We deal with the corrections having $p+q=L$ which we can prove to have a clean ultraviolet origin. 
As a consequence, the exponent $r$ turns out to be bounded as $r\le L$ in this case. 
Our study will also be further limited by the condition $r=L$, {\em i.e.} we shall calculate the leading contributions in the logarithmic expansion.

Since the combined power of factors $m_{t, b}$ is maximal in Eq.~(\ref{eq:yuk}), it is impossible
to generate Yukawa SL in other 1PI parts. For instance, gauge boson self energies will always have too many gauge
couplings. Box diagrams will have suppressed Yukawa couplings to initial state light fermions. 
As in the above examples, similar diagrams occur in other processes like in particular those related to Fig.~(\ref{fig:diagrams}) by supersymmetry, 
as discussed in~\cite{Beccaria:2005xq}.

Due to the counting of couplings, the Yukawa effects  in Eq.~(\ref{eq:yuk}) can be computed in an essentially gaugeless limit of the MSSM.
In particular, the radiative corrections shown in Fig.~(\ref{fig:diagrams}) require the calculation of a  vertex diagram with two 
on-shell chiral and antichiral fields and an external classical vector field, neutral or charged according to the process.

In this gaugeless limit, the only relevant piece of the MSSM lagrangian is the Yukawa superpotential
\be
{\cal W} = y_t\ \overline{t}\ (t H^0_u - b H^+_u) + y_b\ \overline{b}\ (t H^-_d - b H^0_d).
\ee
The chiral fields $\overline t$, $\overline b$ contain the right handed antitop and antibottom, the SU(2) doublet $Q = (t, b)$ is 
composed of the chiral fields $t$ and $b$ containing the left handed top and bottom, the SU(2) doublets $H_u = (H_u^+, H_u^0)$
and $H_d = (H_d^0, H_d^-)$ contain the various Higgs and Higgsino fields of the MSSM.
The couplings $y_{t,b}$ have the tree level value ($g$ is the SU(2) gauge coupling and $M_W$ is the W boson mass)
\be
y_t = \frac{g}{\sqrt{2}}\frac{m_t}{M_W}\frac{1}{\sin\beta} ,\qquad
y_b = \frac{g}{\sqrt{2}}\frac{m_b}{M_W}\frac{1}{\cos\beta} ,
\ee
in terms of the conventional vacuum alignment angle $\beta$. 

About soft breaking terms, we shall see in a moment that the corrections Eq.~(\ref{eq:yuk})
arise as a short distance effect and are thus quite independent on the soft breaking Lagrangian.

After these preliminary discussion and definitions, we 
lift the discussion from the examples of Fig.~(\ref{fig:diagrams}) to the specific universal source of Yukawa enhanced logarithms. 
\begin{figure}[htb]
\vskip 0.7cm
\begin{center}
\leavevmode
\psfig{file=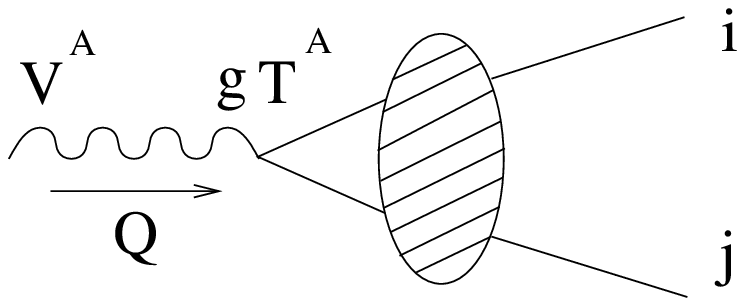,width=6cm,angle=0}
\vspace{0.1cm}
\caption{}
\label{fig:basicdiagram}
\end{center}
\end{figure}
\noindent
This is the basic vertex diagram $V~\to~\Phi_i\overline\Phi_j$ shown in Fig.~(\ref{fig:basicdiagram}), 
where $V$ is a charged or neutral gauge superfield, depending on the specific final state $\Phi_i\overline\Phi_j$.
Let us denote by $g$ the relevant gauge coupling. We want to compute the corrections ${\cal O}(g\ y_t^{2p} y_b^{2q}\ \hbar^{L})$,
$p+q = L$, to the one particle irreducible (1PI) vertex in Fig.~(\ref{fig:basicdiagram}). We denote the
subset of such corrections as $\Gamma^{(V)}_{ij}(Q)$ where $Q$ is a large momentum  entering on the gauge field line
and where the two matter fields are on-shell. Notice that we shall keep working in manifest supersymmetric formalism by employing
superfields. Indeed, perturbation theory will be trivial at the leading logarithmic accuracy with all non trivial effects being 
encoded in the  Renormalization Group (RG) equations.

The fact that $\Gamma^{(V)}_{ij}(Q)$ can be computed in a gaugeless limit with only external classical gauge fields 
and not internal quantum ones is very important and has deep consequences. Indeed, under these conditions,
the large momentum behavior of $\Gamma^{(V)}_{ij}(Q)$ is governed by Renormalization Group (RG) evolution 
equations~\cite{Historical,CollinsEtAl} and does not require the so-called hard evolution equations~\cite{HardEvolution}. 
Technically, in the Callan-Symanzik equation for $\Gamma^{(V)}_{ij}(Q)$ the mass insertion term is irrelevant 
because there are no infrared singularities. As a consequence, one is left with the much simpler
Renormalization Group equation which we can solve. This means that the large logarithms of Yukawa origin are actually
a genuine short distance effect that can be addressed in the deep Euclidean region. 
In particular, they are expected to be independent at leading order on the various mass terms appearing in ${\cal L}_{\rm soft}$.

The RG evolution equation for $\Gamma^{(V)}_{ij}$ is 
\be
\label{eq:rg}
\left(\mu\frac{\partial}{\partial\mu} + \beta_t(y_t, y_b)\frac{\partial}{\partial y_t}
+\beta_b(y_t, y_b)\frac{\partial}{\partial y_b}-\gamma_{ij}(y_t, y_b)\right)
\Gamma^{(V)}_{ij}\left(\frac{Q}{\mu}, y_t, y_b\right) = 0, 
\ee
where $\beta_{t,b}$ are the Yukawa $\beta$-functions, $\gamma_{ij}=\gamma_i+\gamma_j$,  and $\gamma_i$ is  the anomalous dimension of $\Phi_i$.
All RG functions are computed in the gaugeless limit $g\to 0$. The scale $\mu$ is the RG subtraction mass. In deriving Eq.~(\ref{eq:rg}) we 
neglect systematically all non leading terms in the $y_{t,b}$ expansion like gauge boson self energy corrections.

In the next Section, we shall analyze the consequences of Eq.~(\ref{eq:rg})  at the leading logarithmic accuracy.

\section{Renormalization Group analysis}
\label{sec:rg}

As is well known, the general solution of the evolution equation Eq.~(\ref{eq:rg}) is 
\ba
\label{eq:rg2}
\Gamma^{(V)}_{ij}\left(\frac{Q}{\mu}, y_t, y_b\right) &=& \exp
\left\{
   -\int_{\mu}^Q \gamma_{ij}
   \left[
     y_t\left(\frac{\lambda}{\mu}\right), y_b\left(\frac{\lambda}{\mu}\right)
   \right] 
\frac{d\lambda}{\lambda}
\right\}\ \times \nonumber \\
&& \qquad\qquad \times\ \  \Gamma^{(V)}_{ij}\left(1, y_t\left(\frac{Q}{\mu}\right), y_b\left(\frac{Q}{\mu}\right)\right),
\ea
where $y_{t,b}(Q/\mu)$ are the running Yukawa couplings according to the MSSM Yukawa $\beta$-functions.
Eq.~(\ref{eq:rg2}) expresses the renormalized vertex at the scale $Q$ in terms of its value at the initial
scale $\mu$. In principle, we should need the perturbative expansion of $\Gamma^{(V)}_{ij}(1, y)$. However, if we are only interested
in the leading logarithms ${\cal O}(\hbar^L \log^L\frac{s}{\mu^2})$ we can argue from  Eq.~(\ref{eq:rg2}) that they appear in a single 
coefficient multiplying the tree level value of the vertex
\be
\Gamma^{(V)}_{ij}\left(\frac{Q}{\mu}\right) \stackrel{LL}{=} c_{ij}\left(\frac{Q}{\mu}\right)\ \Gamma^{(V), \rm Born}_{ij} ,
\ee
where $c_{ij}$ depends only on the one loop terms in the anomalous dimensions
and $\beta$-functions.

To give an explicit expression for $c_{ij}$ it is convenient to define the coefficients
$\gamma_i^{(1)\ t}$, $\gamma_i^{(1)\ b}$ appearing at one loop in the anomalous dimensions 
($\alpha_{t,b} = y_{t,b}^2/(4\pi)$)
\be
\gamma_i = \gamma_i^{(1)\ t}\frac{\alpha_t}{4\pi} + \gamma_i^{(1)\ b}\frac{\alpha_b}{4\pi} + \cdots,
\ee
as well as the following sums over the external $i$, $j$ fields
\be
\gamma^t_{ij} = \gamma_i^{(1)\ t}+\gamma_j^{(1)\ t},\qquad
\gamma^b_{ij} = \gamma_i^{(1)\ b}+\gamma_j^{(1)\ b}.
\ee
After some algebra, the coefficient $c_{ij}$ turns out to be given by the following compact formula
\be
\label{eq:compactfinal}
c_{ij} = \left[ \frac{\alpha_t(\mu^2)}{\alpha_t(Q^2)}\right]^{\eta^t_{ij}}
\left[\frac{\alpha_b(\mu^2)}{\alpha_b(Q^2)}\right]^{\eta^b_{ij}},
\quad\mbox{with}\left\{\begin{array}{c} 
\eta^{t}_{ij} = \displaystyle\frac{1}{70}(6\gamma^t_{ij} - \gamma^b_{ij}), \\ \\
\eta^{b}_{ij} = \displaystyle\frac{1}{70}(6\gamma^b_{ij} - \gamma^t_{ij}),
\end{array}\right. .
\ee
where the running couplings $\alpha_{t, b}(Q^2)$ are consistently evaluated at leading logarithmic accuracy ({\em i.e.} 
solving the one loop RG equations).

The practical evaluation of Eq.~(\ref{eq:compactfinal}) requires (a) the various anomalous dimensions at one loop, (b) the 
solution of the one-loop RG evolution equations for the running Yukawa couplings. We discuss separately each issue and also provide a perturbative
expansion of  Eq.~(\ref{eq:compactfinal}) whose validity we shall discuss in the applications.

\subsection{Renormalization Group functions}

The RG functions (anomalous dimensions and $\beta$-functions) can be found in the literature~\cite{SuperTwoLoops} for general $N=1$ gauge models.
In the gaugeless limit they reduce at the one loop level to the following simple results that we 
present according to the field numbering $(\Phi_1, \dots, \Phi_8) \equiv (\overline{t}_R, t_L,  H^0_u, b_L,  H^+_u, \overline{b}_R, H^-_d, H^0_d)$.
The first three anomalous dimensions are 
\be
\label{eq:gamma1}
\gamma_1 = 2\left(\frac{y_t}{4\pi}\right)^2,\quad
\gamma_2 = \frac{1}{(4\pi)^2}(y_t^2+y_b^2),\quad
\gamma_3 = \frac{N_C}{(4\pi)^2}\ y_t^2 ,
\ee
The other ones are $\gamma_4=\gamma_2$, $\gamma_5 = \gamma_3$. Also, $\gamma_6$ and $\gamma_7$ are
obtained from $\gamma_1$ and $\gamma_3$ with the replacement $y_t\leftrightarrow y_b$.  Finally $\gamma_8 = \gamma_7$.

The Yukawa coupling $\beta$-functions for the two couplings $y_t$ and $y_b$ are
\be
\label{eq:beta}
\beta_t = \frac{y_t}{(4\pi)^2}
\left[(N_C+3)y_t^2+y_b^2\right],\qquad
\beta_b = \frac{y_b}{(4\pi)^2}
\left[(N_C+3)y_b^2+y_t^2\right].
\ee

\subsection{Running $\alpha_{t,b}(Q^2)$}

From the one loop $\beta$-functions we determine the running Yukawa couplings. In terms of $s = Q^2$ and setting $N_C = 3$
the evolution equations read
\ba
\label{eq:running}
\frac{d\alpha_t}{d\log s} &=& \frac{\alpha_t}{4\pi}(6\alpha_t + \alpha_b), \\
\frac{d\alpha_b}{d\log s} &=& \frac{\alpha_b}{4\pi}(6\alpha_b + \alpha_t), \nonumber
\ea
with given initial conditions at the  scale $\mu$. These equations can be solved analytically in implicit form 
but the exact solution is not particularly enlightening. We provide full details in Appendix~\ref{app:solution}. 
Beside the exact solution, it is interesting to have also its perturbative expansion permitting to discuss the 
convergence of the loop expansion. At second order, we have
\be
\label{eq:leadingrunning}
\frac{\alpha_t(Q^2)}{\alpha_t(\mu^2)} = 1 + \frac{1}{4\pi}(6\alpha_t + \alpha_b)\log\frac{Q^2}{\mu^2} +
\frac{1}{(4\pi)^2} \left(36\alpha_t^2 + \frac{19}{2}\alpha_t\alpha_b+\frac{7}{2}\alpha_b^2\right) \log^2\frac{Q^2}{\mu^2},
\ee
where in the r.h.s. we denote $\alpha_{t, b}\equiv \alpha_{t, b}(\mu^2)$. The result for $\alpha_b(Q^2)/\alpha_b(\mu^2)$
is obtained by exchanging  $\alpha_t\leftrightarrow \alpha_b$. 
The expansion Eq.~(\ref{eq:leadingrunning}) can be substituted in Eq.~(\ref{eq:compactfinal}). Expanding again in powers of the 
logarithms we obtain the leading logarithm approximation for the Sudakov correction $c_{ij}$ (no sum over repeated indices)
\ba
\label{eq:ccomplete}
c_{ij} &=& 1-\frac{1}{2(4\pi)}\log\frac{Q^2}{\mu^2}(\gamma^t_{ij}\alpha_t + \gamma^b_{ij}\alpha_b) + \\
&& + \frac{1}{8(4\pi)^2}\ \log^2\frac{Q^2}{\mu^2}\ [\alpha_t^2\ \gamma^t_{ij}(\gamma^t_{ij}-12) + 
\alpha_b^2\ \gamma^b_{ij}(\gamma^b_{ij}-12) + \nonumber \\
&& \hskip 3cm + 2\alpha_t\alpha_b (\gamma^t_{ij}\gamma^b_{ij}-(\gamma^t_{ij}+\gamma^b_{ij}))] +
{\cal O}(\hbar^3) ,\nonumber
\ea
Eqs.~(\ref{eq:compactfinal}) and its two loop expansion Eqs.~(\ref{eq:ccomplete}) are the main result
of this paper. The one loop approximation (the first line) is in perfect agreement with the results already 
published in the literature~\cite{SudakovMSSM:ILC,SudakovMSSM:LHC1,SudakovMSSM:LHC2} and coming from explicit diagram calculations in component fields followed
by high energy expansions.
In the next Section, we shall discuss in details the consequences of our results by considering 
specific applications.

\section{Applications and Discussion}

\subsection{Preliminary perturbative analysis}

We begin with a discussion of the perturbative result at two loop order  Eq.~(\ref{eq:ccomplete}). This will be useful
in comparing the one loop approximation with higher order corrections.
We denote $L_{t,b} = \frac{\alpha_{t,b}}{4\pi}\ \log\frac{Q^2}{\mu^2}$. From Eq.~(\ref{eq:ccomplete}), we can obtain the 
following list of specific cases
\ba
\label{eq:finalresults}
t_R\ t_R^* &:& c_{11} = 1-2L_t-4L_t^2-L_t L_b , \nonumber \\
t_L\ t_L^* &:& c_{22} = 1-L_t-L_b-\frac{5}{2}(L_t^2+L_b^2) , \nonumber \\
b_L\ b_L^* &:& c_{44} = c_{22}, \nonumber \\
t_L \ b_L^* &:& c_{24} = c_{22}, \\
b_R\ b_R^* &:& c_{55} = 1-2L_b-4L_b^2-L_t L_b , \nonumber \\
H_u^+\ H_u^- &:& c_{55} = 1-3L_t-\frac{9}{2} L_t^2-\frac{3}{2}L_t L_b , \nonumber \\
H_d^+\ H_d^- &:& c_{77} = 1-3L_b-\frac{9}{2} L_b^2-\frac{3}{2}L_t L_b . \nonumber
\ea
The meaning of Eq.~(\ref{eq:finalresults}) is as follows. Let us consider for instance the first line 
which reports the expression of $c_{11}$. This is the correction factor 
for the vertex which has as a final state the chiral field (and its conjugate) whose fermionic
component is the right handed top quark. By supersymmetry, the same correction factor
is also obtained if we consider in the final state various combinations of the scalar partners, for instance
$\widetilde t_R\widetilde t_R^*$. In such a diagonal case, the vector $V$ is necessarily a neutral one, $\gamma$ or $Z$.
Also, by supersymmetry, from the real superfield $V$ we can also take its gaugino component.

Notice that 
the physical charged Higgs boson is a mixture of the charged components of $H_u$ and $H_d$.
A proper two loop result for the vertex with this field would require a one loop treatment of the mixing
that we defer to later work. The vertices with final quark superfields (describing production of 
quark pairs, squark pairs or quark-squark combinations) are correct as they stand, since 
mixing is irrelevant when it appears only in the internal lines.

The above logarithmic expansions Eqs.~(\ref{eq:finalresults}) can be resummed in closed form at least in the single Yukawa coupling limit.
For instance, if we are in a point of MSSM parameters where we can make the approximation $y_b \ll y_t$,
then it is possible to repeat the derivation leading to $c_{ij}$ and exploit the exact solution for 
the running  $\alpha_t(Q^2)$. The final result is remarkably simple
\be
y_b\equiv 0:\qquad c_{ij} = (1-6 L_t)^{\gamma^t_{ij}/12} .
\ee
As an example, in the case of final states $t_R\overline{t}_R$ or $t_L\overline{t}_L$, we find 
(in agreement with the $L_t$ terms in  Eq.~(\ref{eq:finalresults}))
\ba
t_R\ t_R^* &:& c_{11} = (1-6L_t)^{1/3} = 1-2L_t-4L_t^2 + \cdots, \\
t_L\ t_L^* &:& c_{22} = (1-6L_t)^{1/6} = 1-L_t-\frac{5}{2} L_t^2+ \cdots . \nonumber
\ea

\subsection{Discussion of the full result}

Now, we turn to consider the numerical relevance of the computed effects according
to  Eq.~(\ref{eq:compactfinal}) with both Yukawa couplings being active. We concentrate on the same 
processes shown in Fig.~(\ref{fig:diagrams}), i.e. 
\be
\begin{array}{ll}
\mbox{ILC}\qquad & e^+\ e^- \to\ f_\alpha\ \overline{f}_\alpha, \quad f = t, b\\
\mbox{LHC}\qquad & u\ \ \overline{d} \ \ \to\ t_L\ \overline{b}_L. 
\end{array}
\ee
In both cases the polarization of the initial state affects only the tree level amplitude and is factored in the 
correction. The chirality index $\alpha$ determine the polarization of the final state in the ILC process. For brevity,
we consider only a final state with asymptotic vanishing helicity.
The final state of the LHC process has $\alpha=L$ because of $W$ boson exchange.
In both cases, the correction factor in the cross section is simply $(c_{ij})^2$. 

The explicit values of the exponents $\eta^{t,b}_{ij}$ required for the evaluation of Eq.~(\ref{eq:compactfinal})
are summarized in the following table where, for clarity, we replace the pair $(i,j)$ with the fermion in the 
associated chiral field. As we remarked, by supersymmetry the same corrections are obtained with the sfermion components.
\begin{center}
\begin{tabular}{l|cc}
$i$, $j$ & $\eta^t$ & $\eta^b$ \\
\hline
$t_R\ \overline{t}_R\quad$ & $\quad$ 12/35 & -2/35 \\ 
$t_L\ \overline{t}_L$ & 1/7 & 1/7 \\ 
$b_L\ \overline{b}_L$ & 1/7 & 1/7 \\ 
$b_R\ \overline{b}_R$ & -2/35 & 12/35 
\end{tabular}
\end{center}
The initial condition for $\alpha_t(Q^2)$, $\alpha_b(Q^2)$ is fixed as follows. We choose a scale $\mu$ and define at that scale 
\ba
\label{eq:bc}
\left(\frac{y_t}{4\pi}\right)^2 = \frac{\alpha}{8\pi\ s_W^2}\ \frac{m_t^2}{M_W^2}\ (1+\cot^2\beta_{\rm eff}), \\
\left(\frac{y_b}{4\pi}\right)^2 = \frac{\alpha}{8\pi\ s_W^2}\ \frac{m_b^2}{M_W^2}\ (1+\tan^2\beta_{\rm eff}), \nonumber
\ea
where the parameter $\tan\beta_{\rm eff}$ is a scale dependent effective mixing angle.
At tree level, this is the conventional mixing angle. Beyond tree level, it is just a convenient 
parametrization of the initial values of the top and bottom Yukawa couplings. For simplicity, we shall denote in 
the following discussion $\beta\equiv \beta_{\rm eff}$.

Our results are shown in the three Figs.~(\ref{fig:topright}-\ref{fig:bottomright}) for the ILC processes.
The left hand side of each figure shows the (Yukawa) Sudakov correction to the cross section
at one loop and two loop level. We also show a line with the label {\em exact} standing for the 
evaluation of the all order Eq.~(\ref{eq:compactfinal}). The initial condition is obtained from Eq.~(\ref{eq:bc}) with $\tan\beta=40$.
The curves are shown as functions of $\sqrt{s}$ with the RG scale is fixed at the arbitrary value 
$\mu = 100$ GeV, somewhat between $m_b$ and $m_t$. As usual, the choice of the scale $\mu$ cannot be fixed at the leading order
in the logarithmic expansion.

In all the considered cases the two loop correction is practically equivalent to the exact result. Instead, the 
correction to the one loop approximation is appreciable. In the right hand side of the three figures, we show this 
difference as a function of $\sqrt{s}$ for two reference values of $\tan\beta = 2, 40$.

The value of the extra correction is in all cases at the level of 1-2 \%, which can be visible at ILC. The only exception
is the process $e^+e^-\to b_R\ \overline{b}_R$ at small $\tan\beta$ due to the very small value of $y_b$ in that case.

About the considered LHC process, the equality $c_{24}=c_{22}$ implies that the correction factor is identical to that occurring in 
$e^+e^-\to t_L\overline{t}_L$. However, there are important differences deserving a couple of comments. First, the available energy for the 
partonic process will be in general smaller than at ILC reducing the overall effect. Second, it will be non trivial to identify it with the invariant 
mass of the final state $t_L\overline{b}_L$ due to various systematic corrections (for a realistic analysis of 
these effects in $t\overline{t}$ production, see the second reference in~\cite{SudakovMSSM:LHC1}).

\section{Summary and Conclusions}
\label{sec:conclusions}

This paper has been devoted to the analysis of a specific subset of the radiative corrections 
affecting physical processes in the MSSM. In the framework of a high energy expansion we have concentrated 
on large logarithmic corrections to the invariant amplitudes ${\cal A}$ that take the following asymptotic
form at $L$ loops
\ba
{\cal A} &=& {\cal A}^{\rm tree}\left(1+\sum_{L\ge 1}c^{\cal A}_L\ \alpha^L\ \log^L\frac{s}{\mu^2}\right), \\
c^{\cal A}_L &=& \sum_{p=0}^L d_{L, p}^{\cal A}\left(\frac{m_t^2}{M_W^2}\right)^p\left(\frac{m_b^2}{M_W^2}\right)^{L-p} .
\ea
We have shown how to compute the coefficients $\{d^{\cal A}_{L, p}\}$ for various specific processes at ILC or LHC
by solving the Renormalization Group equation governing the relevant 1PI parts from which the above correction originates. The final recipe is quite simple
and universal. It explains various features of the above correction already noted in the literature as a byproduct of 
explicit component calculations. 

In the Standard Model, the coefficients $\{d^{\cal A}_{L, p}\}$ are fixed numerical constants. Instead, in the MSSM, they
can vary by large
amounts as the MSSM parameter space is explored. In benchmark scenarios where one of the two relevant 
Yukawa couplings is large, we can obtain sizable corrections with relevant two loop contributions. The next order
corrections are rather small in all the presented examples. 

We remark that such two loop contributions are difficult to obtain in a straightforward approach that  would
consist in evaluating the relevant Feynman diagrams at generic kinematical points, followed by 
analytic high energy expansions. For instance the two loop contribution is NNLO in the logarithmic expansion
whereas the very powerful tools presented recently in~\cite{Denner:Multiloop} work only at NLO as they stand.
Instead, the calculation of  $\{d^{\cal A}_{L, p}\}$ is quite easy within the framework of the Renormalization Group.

The main point of our paper has been precisely to emphasize that the Yukawa Sudakov corrections can be 
actually computed in a gaugeless limit where they arise not as mass singularities, but as ultraviolet effects.
This observation relies on old results about the asymptotic properties of quantum field theory form factors
and combines them with the modern phenomenological interest toward large Yukawa coupling regions of the 
vast MSSM parameter space. A typical example of such scenario is the 
conventional SPS4 point~\cite{Ghodbane:2002kg}.
Of course, this kind of strategy is well known for static quantities where large logarithms also appear.
A nice example is the recent calculation~\cite{Gambino:2005eh} of gluino lifetime and branching ratios
in Split Supersymmetric models where the large corrections involve $\log(\widetilde{m}/m_{\widetilde{g}})$,
where $\widetilde m$ is the squark and slepton scale and $m_{\widetilde g}$ is the gluino mass. 
In that case, the large logarithms controlled by the top quark Yukawa coupling are also resummed separately.

The aim of our analysis has been that of showing that similar simple RG treatments are possible in the context
of high energy expansions and in particular for 
the Yukawa SL whose existence is somewhat independent of the more involved genuinely Sudakov double
logarithms that require more sophisticated tools. 
From a numerical point of view, our analysis has shown that the higher order SL effect
could be of the percent size at ILC, in which case it would become visible. Less optimistic
conclusions would apply to the considered LHC processes for which the effect appears to be definitely small.
Note, however, that this fact represents in any case a valuable information, telling us that in this sector
the available one loop expansions are certainly sufficient.

Various extensions of the present analysis are possible. For instance, two physically interesting examples at LHC
are $t\overline{t}$ pair production via the partonic processes $gg\to t\overline{t}$, $q\overline{q}\to t\overline{t}$
and the single top quark production mechanisms that we have not discussed, {\em i.e.}
$bu\to td$ and $bg\to tW$. They admit similar corrections localized in specific three point 
vertices~\cite{SudakovMSSM:LHC1,SudakovMSSM:LHC2}
to which the present formalism can be applied to obtain higher order contributions.

\acknowledgments

M. B. wishes to thank F. M. Renard and C. Verzegnassi for many discussions and suggestions 
about the general topic of radiative corrections in the MSSM at high energy. In particular, the
present analysis finds its roots in the results obtained by explicit calculations at one loop 
in the component approach.

\appendix

\section{Solution of the coupled RG equations}
\label{app:solution}

The equations to be solved are 
\ba
\dot\alpha_t(t) &=& \alpha_t(6\alpha_t+\alpha_b), \\
\dot\alpha_b(t) &=& \alpha_b(6\alpha_b+\alpha_t),
\ea
where we have introduced the variable $t = \frac{1}{4\pi}\log(s/s_0)$ that we shall call {\em time} in the following. This dynamical 
system admits the family of integral curves
\be
\label{eq:integral}
(\alpha_t-\alpha_b)^7 = C\ (\alpha_t\alpha_b)^6,
\ee
where $C$ is a constant that can be computed at the initial point $t=t_0$. Exploiting Eq.~(\ref{eq:integral}), 
we can write $\alpha_{t,b}$ parametrically in terms of the ratio $R = \alpha_t/\alpha_b$. 
It is necessary to treat separately the two cases $R(t_0)>1$ and $R(t_0)<1$. If $R(t_0)=1$, then $C=0$ and $R=1$ at all times.
In this case we have simply the single coupling RG solution $\alpha_t = \alpha_b = \alpha_t(0)/(1-7t)$. It is enough to work out
the case $R(t_0)>1$, since the other is obtained by the swap $\alpha_t\leftrightarrow\alpha_b$.
The parametric equations are
\ba
\label{eq:parametric}
\alpha_t &=& C^{-1/5} R^{-1/5}(R-1)^{7/5}, \\
\alpha_b &=& C^{-1/5} R^{-6/5}(R-1)^{7/5}, \nonumber
\ea
where the ratio $R$ satisfies
\be
\label{eq:ratio}
\dot R(t) = 5 C^{-1/5} R^{-1/5}(R-1)^{12/5} .
\ee
We immediately see that $R$ increases as $t$ increases. In practice there is an exploding time at which 
$R\to \infty$. This happens at unphysical large energies for realistic initial values. 
Solving the separable differential equation Eq.~(\ref{eq:ratio}), we find the (implicit) solution for the ratio $R\equiv R(t)$
in terms of its initial value $R_0\equiv R(t_0)$
\be
\label{eq:zero}
F(R)-F(R_0) = C^{-1/5}(t-t_0),
\ee
where the monotonically increasing function $F(R)$ is 
\be
F(R) = -\frac{1}{14}\left[\frac{R^{1/5}(R+1)}{(R-1)^{7/5}}+\frac{1}{3}(R-1)^{3/5}\ {}_2 F_1\left(
\frac{3}{5}, \frac{4}{5}, \frac{8}{5}, 1-R\right)\right].
\ee
Thus, an accurate and quite simple recipe to solve the initial system is the following three steps algorithm: 
(i) compute $C$ from the initial values of $\alpha_{t, b}$, 
(ii) at each time find numerically the unique zero of Eq.~(\ref{eq:zero}), 
(iii) replace the root $R(t)$ in Eqs.~(\ref{eq:parametric}).
This semi-analytic procedure is superior to any discrete step integration scheme, like for instance Runge-Kutta methods.
Indeed, from Eq.~(\ref{eq:zero}) we derive immediately exact properties of the solution. An example is the (unphysical) exploding 
time which is predicted by  Eq.~(\ref{eq:zero}) to be
\be
t_{exp} = t_0+C^{1/5}(F_\infty-F(R_0)), \qquad F_\infty = -\frac{1}{42}\frac{\Gamma(1/5)\Gamma(8/5)}{\Gamma(4/5)}.
\ee

\newpage

\begin{figure}
\centering
\epsfig{file=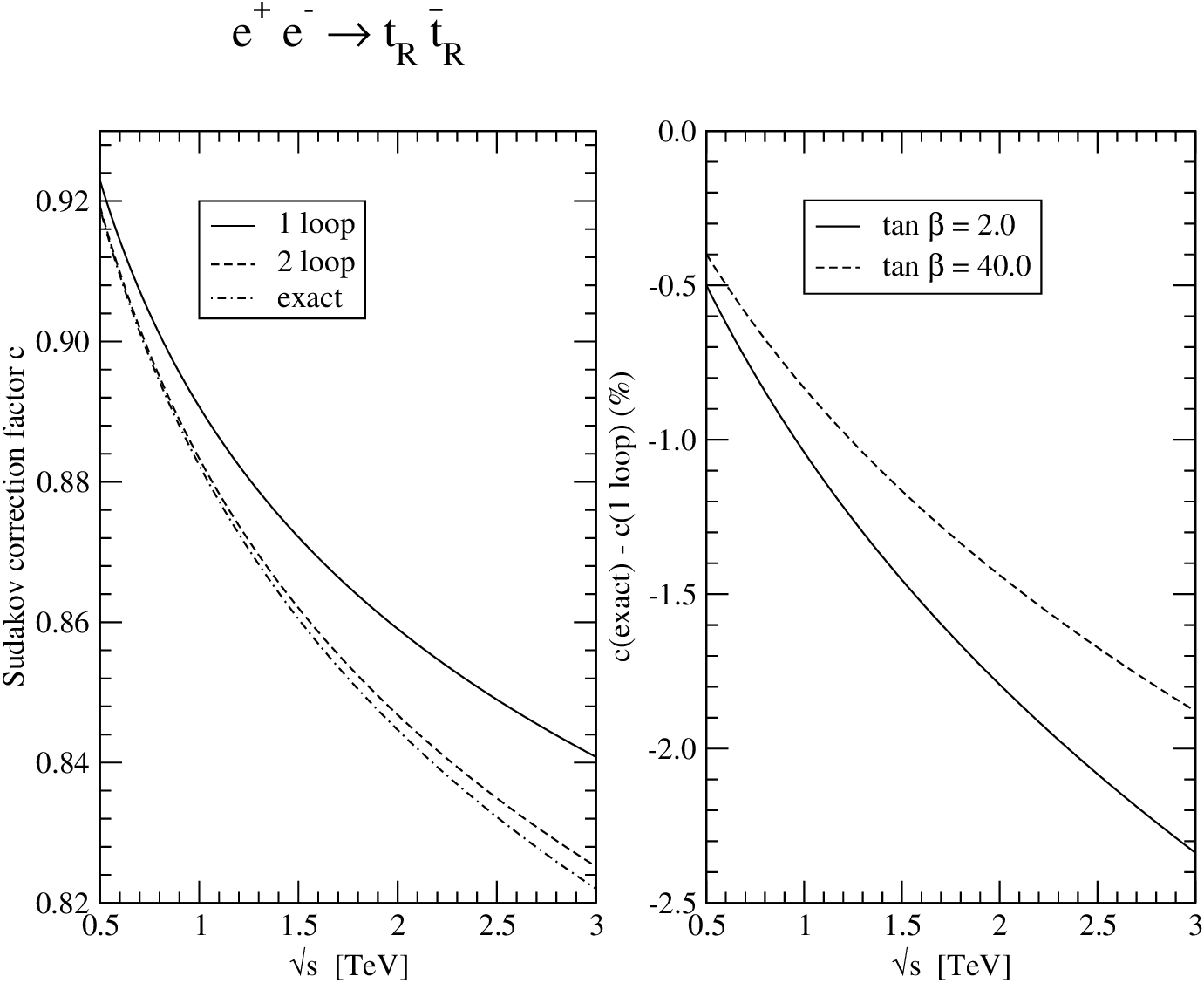, width=18cm, angle=90}
\vspace{1.5cm}
\caption{
Energy and $\tan\beta$ dependence of the Yukawa Sudakov correction factor $c$ in the process $e^+e^-\to t_R\overline{t}_R$.
}
\label{fig:topright}
\end{figure}

\newpage

\begin{figure}
\centering
\epsfig{file=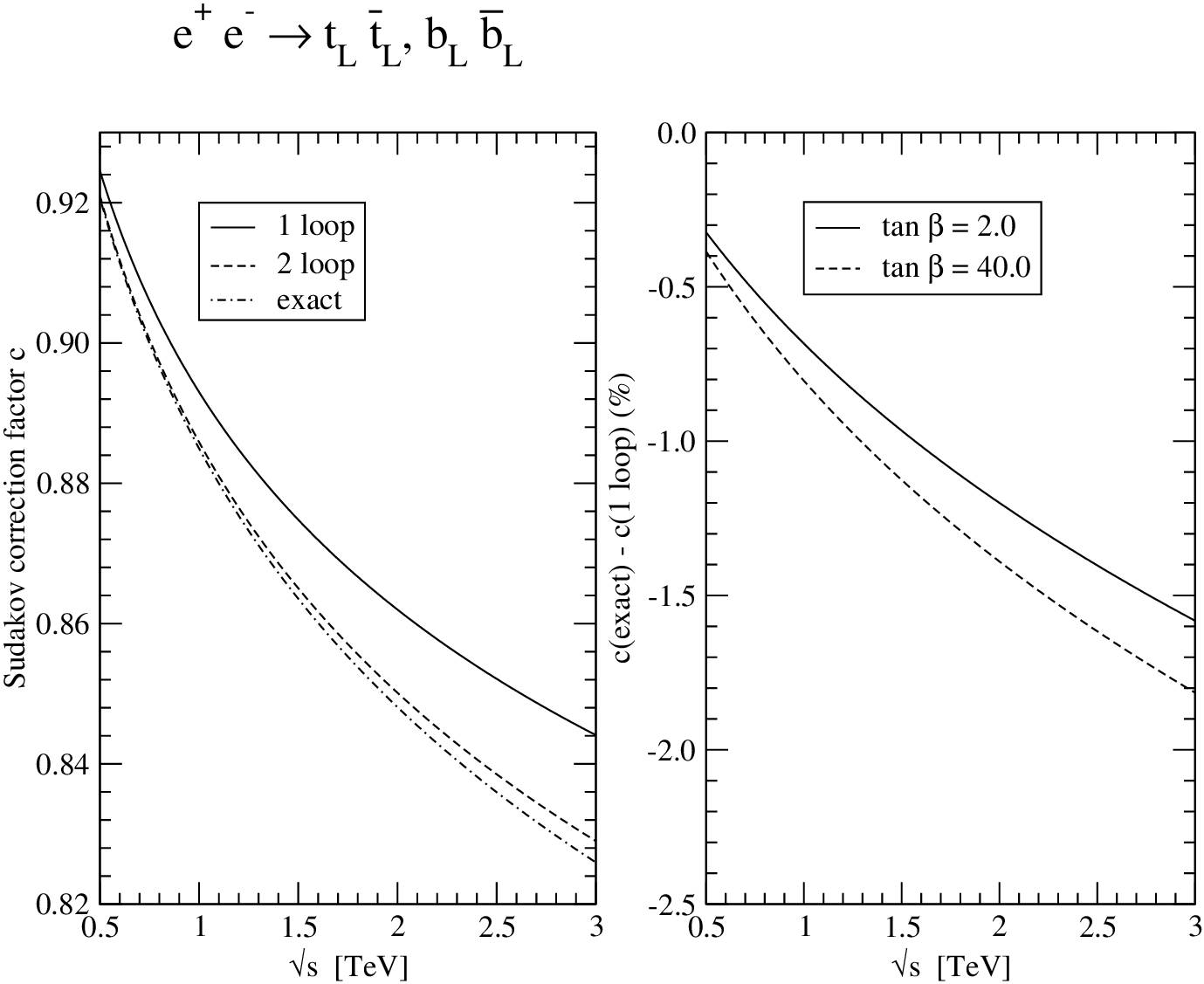, width=18cm, angle=90}
\vspace{1.5cm}
\caption{
Energy and $\tan\beta$ dependence of the Yukawa Sudakov correction factor $c$ in the process $e^+e^-\to t_L\overline{t}_L$
and $e^+e^-\to b_L\overline{b}_L$
}
\label{fig:topleft}
\end{figure}

\newpage

\begin{figure}
\centering
\epsfig{file=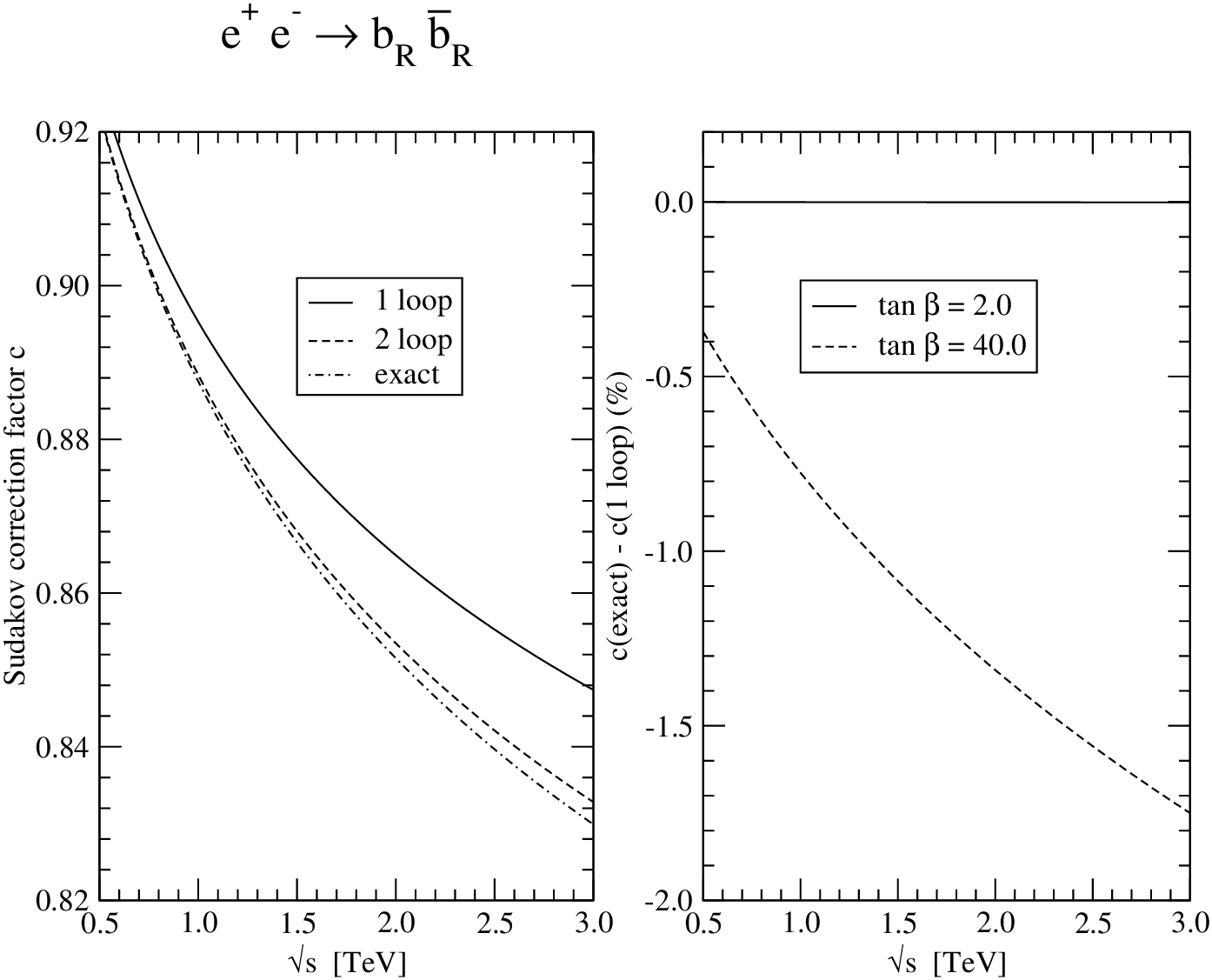, width=18cm, angle=90}
\vspace{1.5cm}
\caption{
Energy and $\tan\beta$ dependence of the Yukawa Sudakov correction factor $c$ in the process $e^+e^-\to b_R\overline{b}_R$.
}
\label{fig:bottomright}
\end{figure}

\end{document}